\documentclass[aps,prl,10pt,twocolumn,showpacs,superscriptaddress]{revtex4-2}
\usepackage{amssymb,graphicx,color,amsmath}
\usepackage{xfrac,bm,stmaryrd,trimclip,soul}
\usepackage[mathletters]{ucs}
\usepackage{gensymb}
\usepackage{siunitx}
\usepackage[colorlinks, linkcolor=blue, anchorcolor=blue, citecolor=blue, urlcolor=blue]{hyperref}

\newcommand{\blue}[1]{ {\color{blue}} }

\newcommand{\SM}{SmB$_6$}

\newcommand{\br}{\mathbf{r}}

\usepackage{lineno}

\begin{document}
\title{Realizing a topological diode effect on the surface of a topological Kondo insulator}

\author{Jiawen Zhang}
\thanks{These authors contributed equally to this work.}
\affiliation  {Center for Correlated Matter and School of Physics, Zhejiang University, Hangzhou 310058, China}

\author{Zhenqi Hua}
\thanks{These authors contributed equally to this work.}
\affiliation  {Department of Physics, Florida State University, Tallahassee, Florida 32306, USA}

\author{Chengwei Wang}
\thanks{These authors contributed equally to this work.}
\affiliation  {Center for Correlated Matter and School of Physics, Zhejiang University, Hangzhou 310058, China}

\author{Michael Smidman}
\affiliation  {Center for Correlated Matter and School of Physics, Zhejiang University, Hangzhou 310058, China}

 \author{David Graf} 
\affiliation  {National High Magnetic Field Laboratory, Tallahassee, Florida 32310, USA}

\author{Sean Thomas}
\affiliation  {Los Alamos National Laboratory, Los Alamos, NM 87545, USA}

\author{Priscila F. S. Rosa}
\affiliation  {Los Alamos National Laboratory, Los Alamos, NM 87545, USA}

\author{Steffen Wirth}
\affiliation  {Max-Planck-Institute for Chemical Physics of Solids, N$\Ddot{o}$thnitzer Str. 40, 01187, Dresden, Germany}

\author{Xi Dai}
\affiliation  {Department of Physics, Hong Kong University of Science and Technology, Clear Water Bay, Hong Kong}

\author{Peng Xiong}
\affiliation  {Department of Physics, Florida State University, Tallahassee, Florida 32306, USA}


\author{Huiqiu Yuan}
\email[Corresponding author: ]{hqyuan@zju.edu.cn}
\affiliation  {Center for Correlated Matter and School of Physics, Zhejiang University, Hangzhou 310058, China}
\affiliation {Institute for Advanced Study in Physics, Zhejiang University, Hangzhou 310058, China}
\affiliation  {Institute of Fundamental and Transdisciplinary Research, Zhejiang University, Hangzhou 310058, China}
\affiliation  {State Key Laboratory of Silicon and Advanced Semiconductor Materials, Zhejiang University, Hangzhou 310058, China}

\author{Xiaoyu Wang} 
\email[Corresponding author: ]{xiaoyuw@magnet.fsu.edu}
\affiliation  {National High Magnetic Field Laboratory, Tallahassee, Florida 32310, USA}

\author{Lin Jiao}
\email[Corresponding author: ]{lin.jiao@zju.edu.cn}
\affiliation  {Center for Correlated Matter and School of Physics, Zhejiang University, Hangzhou 310058, China}

\begin{abstract}\bf
Introducing the concept of topology into material science has sparked a revolution from classic electronic and optoelectronic devices to topological quantum devices. The latter has potential for transferring energy and information with unprecedented efficiency. Here, we demonstrate a topological diode effect on the surface of a three-dimensional material, \SM, a candidate topological Kondo insulator. The diode effect is evidenced by pronounced rectification and photogalvanic effects under electromagnetic modulation and radiation at radio frequency. Our experimental results and modeling suggest that these prominent effects are intimately tied to the spatially inhomogeneous formation of topological surface states (TSS) at the intermediate temperature. This work provides a novel manner of breaking the mirror symmetry (in addition to the inversion symmetry), resulting in the formation of $pn$-junctions between puddles of metallic TSS. 
This effect paves the way for efficient current rectifiers or energy-harvesting devices working down to radio frequency range at low temperature, which could be extended to high temperatures using other topological insulators with large bulk gap.
\end{abstract}

\maketitle

Traditional $pn$-junctions based on semiconductors are prototype devices for realizing nonreciprocal charge transport, which have broad applications in modern electronics, including current rectifiers, photo-detectors, and energy-harvesting systems~\cite{Semiconductor}. However, the exploration of such devices has extended beyond semiconducting heterostructures to bulk homogeneous quantum materials, including ferroelectrics~\cite{Fridkin2001} and Weyl semimetals~ \cite{Osterhoudt2019,SunDongNM2019}. More recently, novel nonreciprocal effects have been proposed in superconducting Josephson junctions and noncentrosymmetric superconductors~\cite{XiDai2007,Pal2022,Lin2022}. In the optical transport regime, $pn$-junction could generate photovoltaic effect. This arises from the nonlinear response of the direct electrical current (dc) $\mathbf{j}(0)$ to an alternating electric field $\mathbf{E}(\omega)$, i.e. $\mathbf{j}_i(0) = \sigma^{(2)}_{ijk}(0,\omega,-\omega)\mathbf{E}_{j}(\omega)\mathbf{E}_{k}(-\omega)$, where $i,j,k=x,y,z$ and $\sigma^{(2)}_{ijk}(0,\omega,-\omega)$ is the second order conductivity tensor \cite{Tokura2018, Nagaosa}. To generate a dc current response, spatial inversion symmetry $\mathcal{P}$ must be broken, which in heterostructures occurs macroscopically near the junction interface, leading to an internal electric field~\cite{Shockley1949}. However for bulk homogeneous materials lacking centrosymmetry, $\mathcal{P}$ is broken on the microscopic scale, and the associated diode effect is better thought of as emerging from the non-trivial topological properties of the Bloch bands. In addition to the aforementioned heterostructures and bulk materials, topological insulators are bulk insulators whose transport is governed by gapless TSS at low temperatures, where $\mathcal{P}$ is naturally broken due to the bulk termination, providing a novel platform for nonreciprocal diode effects~\cite{SCZhang2012}. 
On the other hand, achieving diode effects from TSS requires the breaking of additional mirror symmetries, namely a mirror plane perpendicular to the surface, in addition to $\mathcal{P}$ \cite{Sodemann2015,Kim2017}. Consequently, diode effect is not typically observed in 3DTIs, but have been realized in engineered heterostructures utilizing films of topological materials recently~\cite{McIver2012, C7NR01715D, TaIrTe42021, lu2024nonlinear, kang2019nonlinear, ma2019observation, he2021quantum, dzsaber2021giant, min2023strong, tzschaschel2024nonlinear, Li2015, Eschbach2015, Tu2016}.

\begin{figure*}[ht!!] 
\includegraphics[width=.95\textwidth]{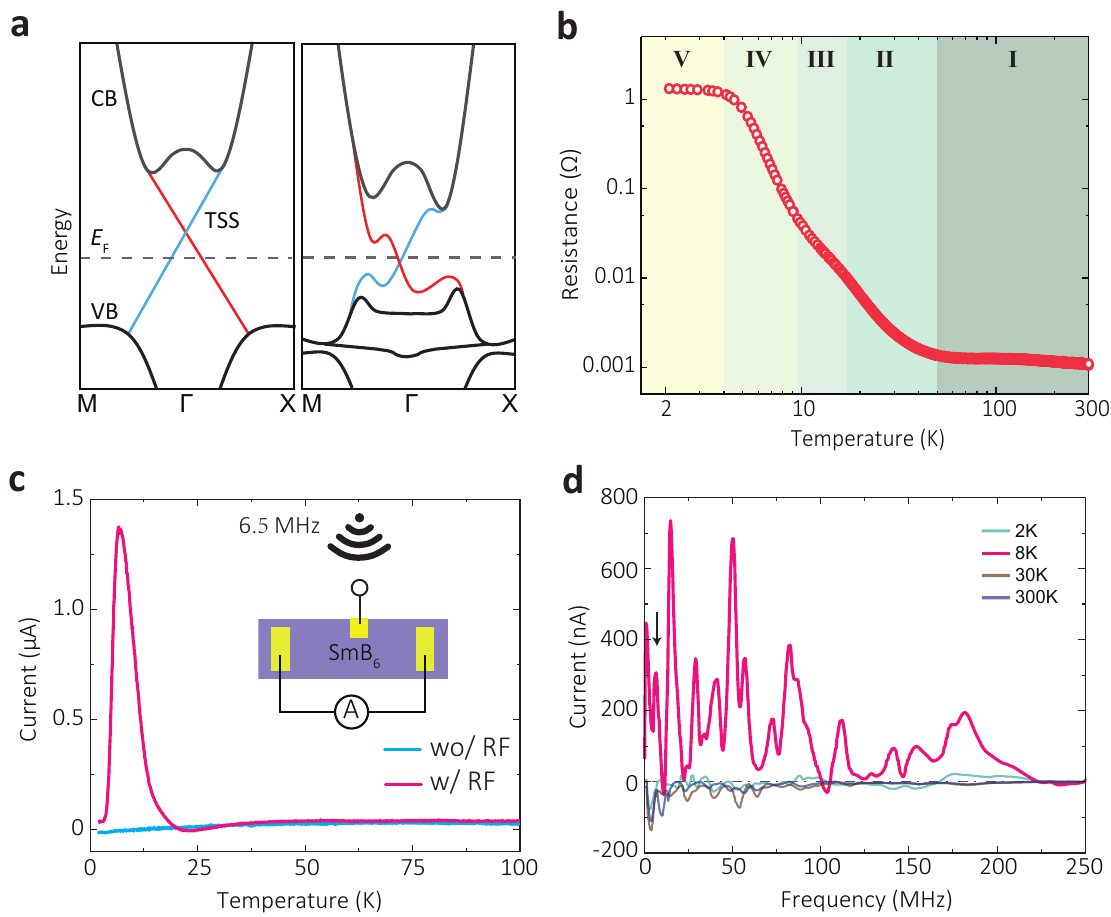}
\caption{\textbf{Rectification effect in \SM.} \textbf{a}, The band structure of an ideal weakly correlated topological insulator (left), and \SM\ (right). The bulk gap in \SM\ is very small and the Dirac cone is heavily renormalized. \textbf{b}, Typical temperature dependent resistance $R(T)$ of \SM\ on a log-log plot. The data was obtained by the standard four-contact method. The color bands mark five different regions according to the slope of $R(T)$. \textbf{c}, Rectification current measured by the two-contact method. A $-$20~dBm 6.5~MHz RF signal was applied to the sample through a third electrode. A large positive current peak emerges at around 7~K. The blue curve is a measurement of the same contacts without applying the RF signal. The inset displays the experimental setup. \textbf{d}, Frequency response of the rectification current at selected temperatures. The signal power was $-$20~dBm for all the temperatures except 300~K, at which $-$12~dBm signal was applied. The black arrow indicates the 6.5~MHz frequency, which is mostly used in this work.}
\label{fig1}
\end{figure*}

\SM\ is a potential 3DTI driven by strong electronic interactions~\cite{Coleman2016,Allen2020}, for which there is compelling evidence of topologically protected surface states that fully develop below around 4~K, stemming from inversion of the $f$- and $d$-bands via the Kondo effect~\cite{Kim2013,Fisk2019,Paglione2013,Hoffman2020}[Fig.~\ref{fig1}(a)].
Unlike typical 3DTIs~\cite{RMP}, the much smaller bulk gap of around 3-5~meV~\cite{Kunii2001,Flachbart,ChenXHPRB,Jiao2018} leads to the resulting TSS being susceptible to various perturbations such as defects, strain and surface reconstructions~\cite{wirth2021a}. Both experiment and theory have discussed heavy and light Dirac electrons coexist on the surface of \SM~\cite{ColemanPRL2016,Kawakami2016,Jiao2018}. Such unique characteristics are proposed to account for many exotic experimental observations in \SM\ and highlight the potential of the surface states for practical applications~\cite{XiaJing, Jurgen2019, Takeuchi2019a, Madhavan2022}. 

In reality, electrons on the surface of \SM\ are inevitably subjected to various defects, stress, and surface reconstructions~\cite{Fisk2014, Jiao2018, SunZX, Pagliuso2020}. Given the tiny energy scales associated with strong correlations, surface perturbations offer a natural way to reduce the spatial symmetry: while some regions may have gapless TSS, other regions might remain gapped, and some may exhibit lighter surface states due to surface Kondo breakdown~\cite{ErtenPRL2015}. 
As the spatial inhomogeneities break mirror symmetries, the diode effect is expected at the boundary between p- and n-type TSS.
In this study, we demonstrate that \SM\ exhibits a pronounced rectifying dc current under radio frequency (RF) modulation. The rectification effect intimately ties to the (partial) formation of TSS, which forms self-generated $pn$-junction. Such junction could efficiently covert RF radiations into DC power.   

\begin{figure*}[ht!!] 
\includegraphics[width=.95\textwidth]{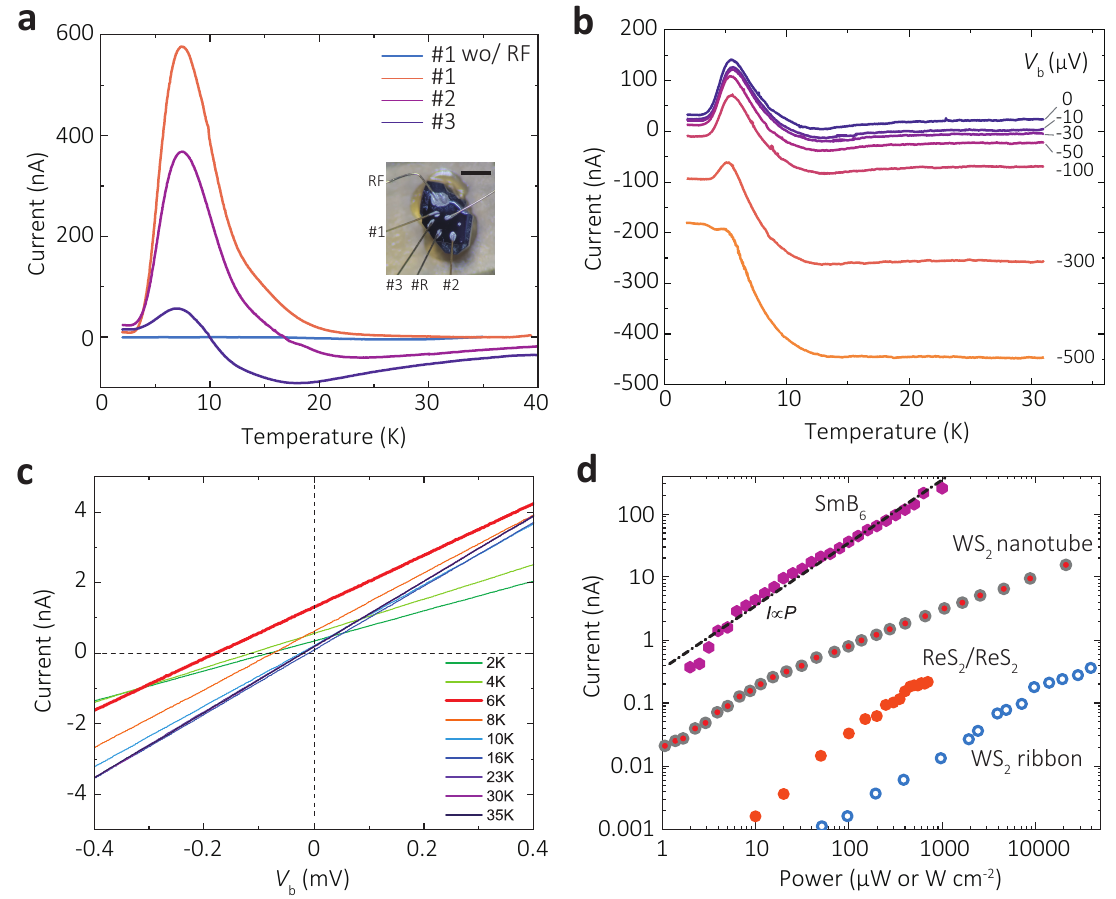}
\caption{\textbf{Performance of the $pn$-junction.} \textbf{a}, Rectification current between three contacts and a reference contact (labeled as \#R) on polished \SM. The current peak value varies significantly and is distributed randomly across the surface. Inset: optical photograph of the sample. The scale bar is 1~mm. \textbf{b}, Current measured while various bias voltages ($V_b$) are applied across the two-probe circuit. \textbf{c}, \textit{IV}-curves measured by placing two electrodes
on sample S1 between 2 K and 35 K. Large photogalvanic current between 5 to 15~K is observed as demonstrated by the intersect of the \textit{IV}-curves at zero bias. \textbf{d}, Power-dependent photogalvanic currents of \SM, which is compared with the ``shift current" observed in three low-dimensional transition metal dichalcogenides (data is adopted from Ref.~\cite{zhangNature,LiangNC,XueScience}). The unit of $x$-axis is $\mu$W for \SM\ and edge-embedded ReS$_2$, while the unit is W cm$^{-2}$ for WS$_2$ nanotube and ribbon.}
\label{fig2}
\end{figure*}

Figure~\ref{fig1}\textbf{b} illustrates the typical temperature-dependent resistance $R(T)$ of \SM, measured using a standard four-probe configuration. The $R(T)$ curve is divided into five regions based on the slope of the log-log plot. In region I at high temperatures covering the widest temperature range, \SM\ behaves as a correlated bad metal. An exponential increase is observed in region II upon cooling as the Kondo hybridization gap opens~\cite{Geb1969}. Notably, $R(T)$ displays two inflection points near 15~K and 10~K, defining the boundaries for regions III and IV. These temperature scales reflect the crossover region where TSS begin to play a role~\cite{Fisk2014,Wu2018,Fisk2019}. For instance, narrow in-gap states emerge between 15~K to 5~K in point-contact spectroscopy measurements~\cite{Flachbart, Park};
magnetoresistance and its rotational symmetry changes imply a transition from bulk to surface state dominated transport between 16~K and 5~K~\cite{ChenXHPRB}; and helical tunneling sets in around 10~K~\cite{Madhavan2022}. Thus, these regions indicate a crossover to where surface effects overtake the bulk contributions~\cite{Rakoski}. When region V is reached, $R(T)$ plateaus, indicating that the TSS fully dominate the electrical conductivity.

\begin{figure*}[ht!!] 
\includegraphics[width=.95\textwidth]{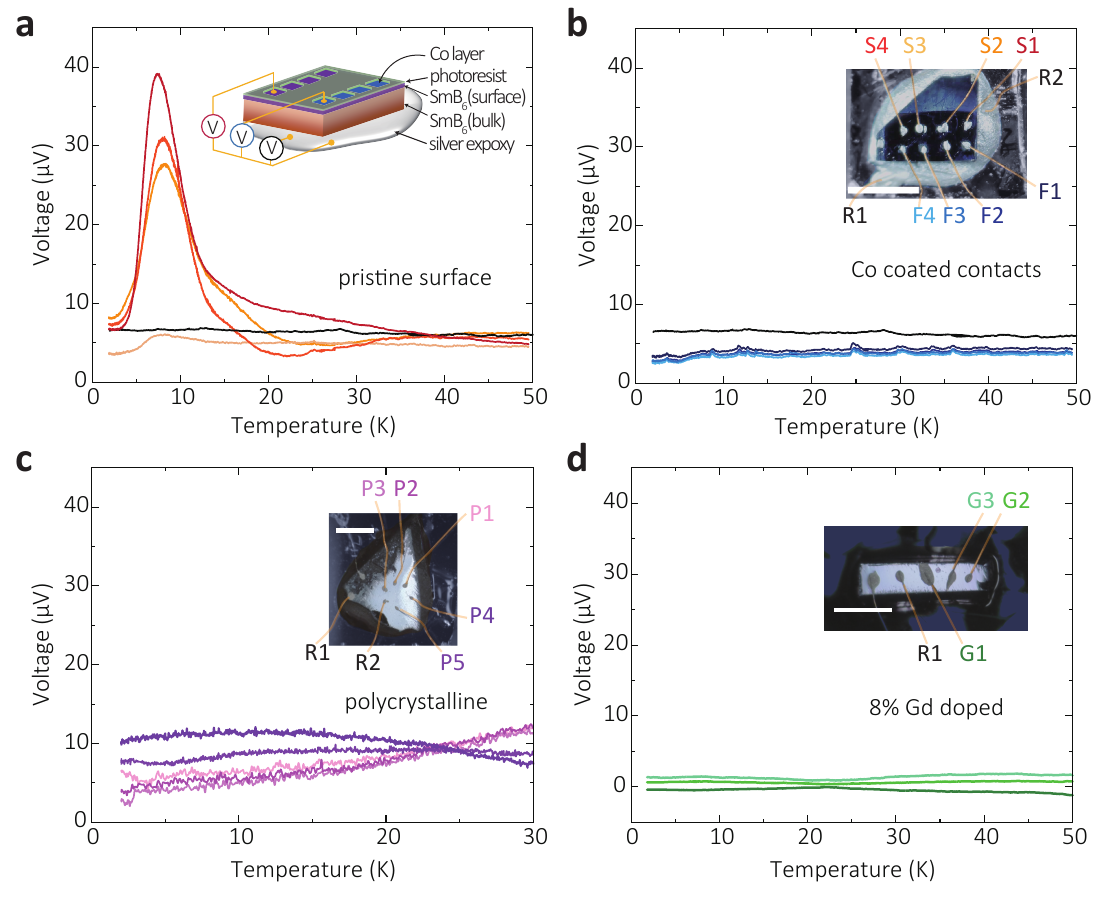}
\caption{\textbf{Tuning the TSS and the surface rectification effect.} \textbf{a}, Rectification voltages between different pristine surface contacts and the reference contact R1. The curves are color coded to correspond to the contact labels in the inset of figure~\textbf{b}. The black line is the voltage measured between the two reference electrodes on silver epoxy. Inset: schematic of the sample and experimental setup used for obtaining the data in \textbf{a} and \textbf{b}. \textbf{b}, Rectification voltages between the four Co-coated contacts and R1. Inset: optical photo of the sample (S3) with pristine surface contacts (labeled S1 to S4) and Co-coated contacts (labeled F1 to F4). \textbf{c}, $V(T)$ of  polished polycrystalline \SM. The voltages of P1, P2, and P3 are referred to R1, while for P4 and P5 R2 was the reference. Inset: optical photo of the sample. \textbf{d}, $V(T)$ of an 8\% Gd doped sample. The voltages were measured between G1, G2, and G3 and the reference R1. Inset: optical photo of the sample. All the white scale bars represent 2~mm in the photographs. An RF signal with a frequency of few MHz was applied to the sample directly via contact wires.}
\label{fig3}
\end{figure*}

In the following, we focus on the emerging surface state in the intermediate temperature regime (regions III and IV). We note that the standard 4-probe method using alternative current (ac) eliminates any unidirectional potential.
To study the diode effect, we measure the current or voltage between two electrodes positioned either on cleaved or carefully polished surfaces. The details regarding sample preparation and measuring techniques can be found in the Supplementary Information (SI) Methods section and Figs. S1 to S2.
In most cases, the electrical current ($I(T)$) is derived by measuring the voltage drop across a 100~$\Omega$ resistor in the circuit (see Fig. S2). Without adding a RF modulation on one electrode or applying RF radiation, we only detect the trivial thermoelectric/contact voltage as a tiny (typically $\leq$10 $\mu$V) and nearly temperature-independent background, see the blue line in Fig.~\ref{fig1}\textbf{c} and Fig. S3.

Interestingly, upon being exposed to a RF signal (e.g. at a frequency of 6.5~MHz), a pronounced dc current signal develops below 20~K and peaks around 7~K. 
One might reasonably assume that a $pn$-junction(s) or Schottky barriers induce such a rectification current. 
Before discussing further, it is important to consider a scenario where a Schottky barrier forms between \SM\ and the electrodes. A characteristic behavior of a Schottky barrier is a non-linear $IV$-curve due to non-ohmic contacts. However, when \SM\ is exposed to a RF signal, the $IV$-curve maintains linear behavior (see Figs.~\ref{fig2} and S4) while a nonzero intersection at zero bias voltage appears between 5-15~K. Moreover, the rectification effect, despite the different maximum value in DC current, is always visible across different electric contacts, including three different types of silver epoxy/paint, and gold-coated electrodes. Given that the Schottky barrier is dependent on the work function of the constituent materials, it should be extremely sensitive to the nature of the metallic electrodes at the sub-meV energy level. Therefore, it is unlikely that the Schottky barrier/diode is the mechanism causing the observed large rectification effect. 

To better understand the diode effect observed in \SM, we measured the RF frequency response of the DC current at different temperatures ($I(T)$). In the temperature regimes I, II, and V, i.e. 300~K, 30~K, and 2~K, only a small dc signal is detected, which is considered as a trivial background signal (see Fig.~\ref{fig1}\textbf{d}). These three temperatures correspond to  where \SM\ is a well-defined bad metal, a Kondo insulator, and a topological Kondo insulator with fully developed TSS, respectively. In these cases, no intrinsic rectification effect is expected on the surface.
On the other hand, in the intermediate temperature regime (8~K), the current exhibits a strong frequency response, suggesting that the diode effect appears in the temperature regime where TSS are developing. 

The observation of multiple peaks at different frequencies (red curve in Fig.~\ref{fig1}\textbf{d}) indicates a complex configuration/structure of the self-generated $pn$-junctions. Other possible origins could be the complicated band structure of \SM\ due to crystal electric field~\cite{Dai2013} or defect induced band bending. Interestingly, our STM measurements have visualized spontaneous inhomogeneity in topography only at such an intermediate temperature, see Fig. S5.

To investigate the spatial distribution of the self-generated junctions, we randomly positioned several electrodes on a freshly polished sample (S2, inset to Fig.~\ref{fig2}\textbf{a}) and measured the rectifying current/voltage between the reference electrode (consistently denoted as '\#R') and the other three electrodes. The peak amplitude at $\sim$7~K of $I(T)$, Fig.~\ref{fig2}\textbf{a}, or $V(T)$ shows significant variations of rectification effect between electrodes, although no clear pattern can be discerned.
In Fig. S6, we reproduce the observation on two additional samples. These findings imply the diode effect arising from the spatial inhomogeneity of the developing surface states.
Another indication is the fact that below $\sim$4~K, the diode effect has been suppressed in all cases. As the TSS have fully developed and formed a metallic surface at in region V~\cite{Fisk2014}, and spontaneous spatial inhomogeneity should be suppressed away from topological phase transition temperature. Moreover, the formation of insulating gapped regions between p- and n-type TSS should play an important role in pronounced diode effect, which is included in the discussion part.
Further evidence comes from the dependence of $I(T)$ on the bias voltage ($V_b$), where $V_b$ ranges between +3~mV to $-$3~mV (see Fig.~\ref{fig2}(b)). In this scenario, a small $V_b$ is superimposed on the two-probe circuit. Above 15~K, the $I(T)$ curves remain relatively flat for different $V_b$ values because the small two-probe resistance of \SM\ (dominated by the contact resistance) is temperature insensitive in this range (see Fig.~\ref{fig1}\textbf{b} and Fig. S7). The offset of the $I(T)$ curves along the $y$-axis simply follows Ohm's law ($I \propto V_b$). Notably, the rectification current (peaked around 7~K) is only pronounced when $|V_b|<0.3~\mathrm{mV}$ and decreases with increasing $|V_b|$. Above $\pm3~\mathrm{mV}$, the current peak is nearly indistinguishable, and $I(T)$ reverts to the expected $V_b/R(T)$ profile (see Fig. S6 for detailed analysis).
Based on Fig.~\ref{fig1}\textbf{d} and Fig. S7, we infer that the bulk gap of \SM\ likely lies between $\sim$1.5 and $-3~\mathrm{mV}$. This tiny Kondo hybridization gap aligns with the measured thermally activated gap (3-5~meV)~\cite{Kunii2001,Flachbart,ChenXHPRB,Jiao2018} and indicates the in-gap surface states contribute to the additional DC current. We further consolidate our understanding by performing \textit{IV}-curve measurements in Fig.~\ref{fig2}\textbf{c}. Again, a clear finite zero-bias current is observed between 5-15~K (see full dataset in Fig. S4), which resembles ``shift current" observed in many other optoelectronic devices. Therefore, we compare the performance (output power (density) of radiation generator ($P$) versus Short-circuit  DC current) of \SM\ with three recently engineered transition metal dichalcogenides~\cite{zhangNature,LiangNC,XueScience}. For \SM, $I(P)$ shows a linear behavior over 3 orders as expected. Most importantly, for a relatively small input RF radiation, we obtained more than 100~nA DC current through one junction. 
Even if the RF signal is as weak as $-$23 dBm, \SM\ still generates a measurable photo-current. Although a more realistic evaluation of the efficiency relies on a precise determination/comparison of current density, which is inaccessible in our case, the working frequency and power are significantly lower than many other low-dimensional topological materials~\cite{McIver2012, C7NR01715D, TaIrTe42021, lu2024nonlinear, kang2019nonlinear, ma2019observation, he2021quantum, dzsaber2021giant, min2023strong, tzschaschel2024nonlinear}.
We note that the wavelength of the used radio frequency is as large as 100~meters and our sample-antenna distance is around a few millimeters, therefore, our setup is well below the near-field limit and the actual radiant energy on the sample is near zero. This makes \SM\ a promising material for designing photo-detectors or for harvesting energy from thermal radiation and other environmental electromagnetic waves.

\begin{figure*}[ht!!] 
    \centering
    \includegraphics[width=0.95\linewidth]{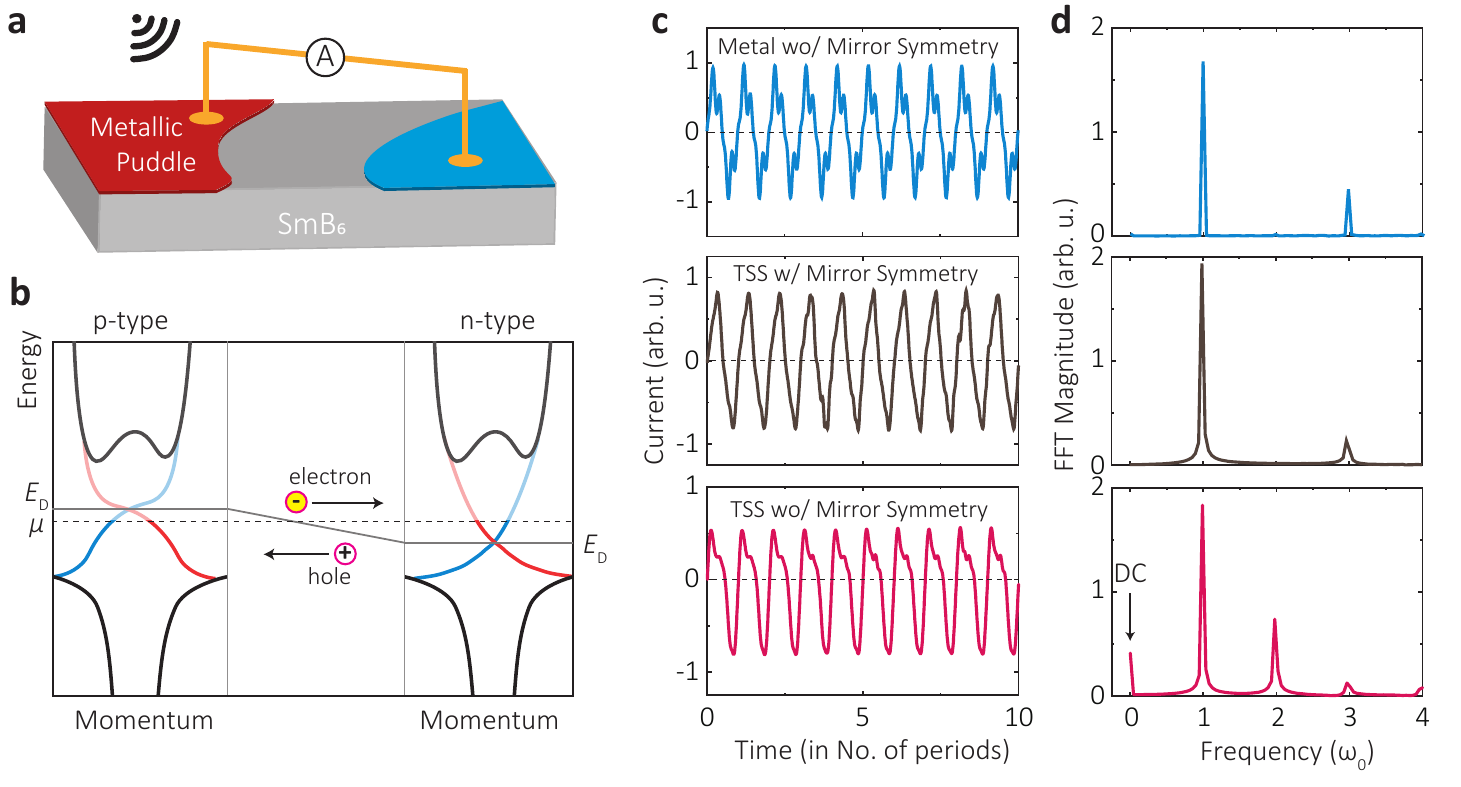}
    \caption{\textbf{Toy model for analyzing the self-generated topological $pn$-junction.} \textbf{a}, Illustration of charge puddles forming out of the insulating \SM\ bulk states. Rectification current is detected by shunting two puddles with a closed circuit loop. \textbf{b}, A putative picture for the formation of a $pn$-junction between different puddles on the surface, which is subjected to RF radiation. Simplified band structures of \SM\ on different puddles are plotted on the left and right columns. Excited electrons (holes) move to the $n$-type ($p$-type) region to form a persistent current. $\mu$ represents chemical potential of \SM\ and $E_D$ marks the energy level of the Dirac point. \textbf{c}, Numerical calculation of the longitudinal current response across junctions between trivial metal surfaces without mirror symmetry (upper panel), TSS with mirror symmetry (middle panel), and TSS without mirror symmetry (lower panel), respectively.  The modeling considers an ac voltage with frequency $\omega_0=0.05t\  \rm rad/sec$ applied along the yellow wire in the circuit, as sketched in \textbf{a}. \textbf{d}, Fast Fourier transform (FFT) of the current signals in \textbf{c}.}
    \label{fig4}
\end{figure*}

Previous research has shown that the TSS in 3DTI can be disrupted by proximity coupling to magnetic layers~\cite{Samarth2013}. Here we further used photolithography to fabricate eight electrode windows on sample S3 (see inset to Fig.~\ref{fig3}\textbf{a}). Four of these were coated with a 30~nm-thick Co layer, while the remaining four had pristine \SM\ surfaces.
We labeled the four Co contacts as F1 through F4, the direct contacts to the pristine \SM\ surface as S1 through S4, and the two contacts on the bottom silver epoxy as R1 and R2.
Figure~\ref{fig3}\textbf{a} shows the measurements of the dc voltage taken between the pristine surface contacts and the reference contact R1. These results align with those presented in Fig.~\ref{fig2}(a) and demonstrate that all four contacts exhibit a maximum rectification effect around 8~K with varying magnitudes. Moreover, the reference channel signal (illustrated by the black lines in Fig.~\ref{fig3}\textbf{a} and \textbf{b}) remains flat and featureless. Interestingly, as shown in Fig.~\ref{fig3}\textbf{b}, all $V(T)$ readings from the ferromagnetic Co contacts did not show any temperature dependence, closely resembling the black reference curve. It is important to note that the Co coating is confined to the contact windows, ensuring the preservation of TSS between Co contacts.

Consistent with previous studies~\cite{ChenYL2010,Hasan2012,Davis2015}, the time-reversal-symmetry protected TSS are not robust to magnetic impurities. For comparison, we conducted measurements on 8\% Gd-doped \SM\ crystals as well as polycrystalline \SM. In both cases the four-probe resistance measurements reveal a similar bulk gap to pristine \SM\ but there are no signatures of resistance saturation down to 2~K (as illustrated in Fig. S8). Such behavior suggests the presence of a fully-opened bulk gap without TSS. Remarkably, neither the polycrystalline \SM\ nor 8\% Gd-doped \SM\ crystals exhibit a measurable rectification effect, as shown in Figs.~\ref{fig3}\textbf{c} and \textbf{d}. Both observations demonstrated the bulk and surface quality could largely affect the performance of the $pn$-junctions. One could further make use of such impacts to design integrate diodes by making patterns of magnetic impurities doped areas. On the other hand, this comparison further rules out Schottky barriers as the cause of the diode effect, since both the pristine and the doped \SM\ display a similar bulk gap.

The experimental results presented above provide signatures that the current rectification observed in \SM\ is linked to the (incomplete) formation of TSS. It is important to note that spatially homogeneous TSS described by the spin-orbit coupled Dirac Hamiltonian do {\it not} lead to current rectification if mirror symmetries along the [100] and [010] directions are preserved, which prohibits both the Berry curvature dipole~\cite{Sodemann2015} and the shift current~\cite{Kim2017} contributions to the nonlinear response. However, the formation of TSS might be spatially inhomogeneous in the intermediate temperature range of 5~K to 15~K, which naturally breaks the mirror symmetry on the mesoscopic scale, as illustrated in Fig.~\ref{fig4}(a). Inhomogeneity strongly modifies the energy dispersion of the TSS, which have been shown to be highly sensitive to the local environment, particularly the valence of ${\rm Sm}$~\cite{ErtenPRL2015}. Such inhomogeneity or puddles have been theoretically predicted~\cite{Skinner2013} and experimentally observed in prototype topological insulators, such as pristine or doped Bi$_2$Te$_3$~\cite{Bathon2016, Rischau2016, Yazdani2011} and BiSbTeSe$_2$~\cite{Knispel2017}. The size of the observed puddles ranges from tens to hundreds of nanometers~\cite{Rischau2016,Yazdani2011,Knispel2017}. Upon further decreasing temperature, the TSS percolate and become more homogeneous. Different experimental probes revealed that TSS dominate transport properties below 3-4 K~\cite{Kim2013, Fisk2019}. This is not a surprise as local phase separation is normal around the phase transition temperature, while electrons tend to condense into a uniformed phase at low temperature. This results in restored mirror symmetry and leads to suppression of the current rectification signal as we observed here.

To provide a quantitative basis for the above picture, we performed a numerical calculation of current rectification due to spatially inhomogeneous TSS. This is achieved by adding to the spin-orbit coupled Hamiltonian terms that depend on the spatial coordinates, $H(\br) = v_F(\sigma_xp_y-\sigma_yp_x)+V(\br)\sigma_0 + m(\br)\sigma_z$. Here $(\sigma_x,\sigma_y,\sigma_z)$ are spin Pauli matrices, $v_F$ is the Fermi velocity, $V(\br)$ is the charge potential that does not open an energy gap to the Dirac dispersion, and $m(\br)$ is the Dirac mass term which can be generated for instance by proximitizing TSS to a magnetic domain. Both $V(\br)$ and $m(\br)$ can tune the mirror symmetry of the system. For example, having $V(x,y)\neq V(-x,y)$ breaks the mirror symmetry with respect to the $yz$-plane. Above continuum single-electronic Hamiltonian is regularized on a square lattice (not to be confused with the atomic lattice), and the system's dynamical transport properties when $V(\br)$ and $m(\br)$ are varied are studied using the time-dependent quantum transport package Tkwant \cite{Kloss_2021}. Details of the lattice regularization and simulation results are presented in SI Methods section. The longitudinal current response ($I_x$) in the presence of a driving ac electric field ($E_x$) are presented in Figs.~\ref{fig4}\textbf{c} and \textbf{d}. An ac electric field indeed lead to a rectifying current along $x$ direction (marked DC) when the mirror symmetry is broken ($V(x,y)\neq V(-x,y)$), but only leads to higher harmonic signals when the mirror symmetry is preserved. For comparison, we also included calculations for the case that the central region is trivially metallic, which leads to negligible current rectification effects. Figure S9 shows more detailed simulation results when a Dirac mass term is added, yielding similar results to those presented here.

In conclusion, we demonstrated that \SM\ displays a significant current rectification (as well as photogalvanic) effect across an intermediate temperature range. This effect, as confirmed both experimentally and through modeling, is ascribed to the broken mirror symmetry associated with the uneven development of TSS. The puddles of primitive TSS serve as reservoirs of p- and n-type Dirac electrons. Our study thus provides an example of a nonreciprocal diode effect in a topological $pn$-junction based on a bulk 3DTI, and offers physical insights into the formation of TSS following the Kondo gap opening. These insights suggest a potential application of 3DTI by introducing spatial inhomogeneity. For instance, the RF frequency response of \SM\ could be utilized for a micro-scale signal detector in the RF range. Although it is limited to very low temperature for this compound, our paradigm could be extended to other 3DTIs with much larger bulk gap. 
It is imperative to note that the typical size between p- and n-type TSS regions should be comparable to the distance between electrodes in order to achieve a substantial current rectification, as otherwise the diode effect may be averaged out. However, this is hard to be realized for nano- or even micro-sized puddles. One practical way is to introduce insulating gapped regions between dilute puddles above 4~K. As presented in Fig. S9 the formation of gapped regions in the intermediate temperatures do not disrupt diode effect, which could ensure separated p- and n-type TSS, and further reduced the averaging effect up to millimeter scale. Certainly such benefit fades away at low temperatures when the TSS percolates and/or become homogeneous. Future experiments employing scanning probe microscopy could further validate the evolution of the TSS puddles on the surface within the intermediate temperature window. Regarding practical applications, these gapped regions could be realized by the controlled introduction of magnetic dopants confined to a specific surface pattern.

\section*{Materials and methods}
\textbf{Material preparations:}

\SM\ single crystals were synthesized by the Al-flux method. Polycrystalline \SM\ with a nominal purity of 99.5\% was produced by Xi'an Function Material Group CO., LTD. Voltage or current was measured on either as-grown, cleaved or polished surfaces, while the latter has much-improved reproducibility and manipulability. The following steps provide recipes for sample polishing:

1. Fix \SM\ crystal on a glass slide or silicon wafer with bonding material (e.g. GE Varnish, PELCO$^\circledR$ Crystalbond\texttrademark\ 509/590, LOCTITE$^\circledR$ Stycast 2850, or EPO-TEK$^\circledR$ H21D silver epoxy). The desired surface should be parallel to the substrate. In most of the cases, we polish the (001) surface.

2. Polishing \SM\ starts with 30 $\mu$m Allied$^\circledR$ aluminum oxide lapping film. We use isopropanol as a lubricant. After every half minute of polishing, we clean the crystal surface and lapping film using clean dry nitrogen. It is important to polish the surface evenly, so there is no mismatch to the desired surface.

3. Repeat step 2 with different sizes of lapping film (12 $\mu$m, 9 $\mu$m, 3 $\mu$m, 1 $\mu$m, 0.3 $\mu$m, and 0.05 $\mu$m), from larger size to smaller abrasive grain size.

4. To polish multiple surfaces on one piece of crystal, Crystalbond\texttrademark\ 509/590 is used to fix the sample. After polishing one surface, we are able to remove the crystal by heating the adhesive, and then remove the residue with the designated solvent. The roughness of the polished surface is checked by atomic force microscopy (AFM) as shown in Fig. S1

Soon after polishing, several electrical contacts were placed on the surface with 15~$\mu$m platinum wires. DuPont\texttrademark\ silver paint was used as an adhesive. The prepared sample is then loaded into a Physical Property Measurement System (PPMS) or a 4 K cryostat from Janis for temperature dependent surface potential measurements. Figure S2 demonstrates the measuring setups. Current was measured by two methods. A simple solution is to short two electrodes by a resistor (e.g. $\sim$100 $\Omega$) and we detect the voltage drop across the resistor. On the other hand, the current signal could also be converted into voltage by a Femto$^\circledR$ current amplifier (DLPCA-200). The voltage signal was measured by either Keithley 2182A Nanovoltmeter or Keithley 2400 SourceMeter$^\circledR$. To precisely measure rectification effect, two 1~KHz low-pass filters are added to the two probes, in order to remove high frequency noise. The radio frequency modulation/radiation are applied to the sample through copper wire either connected to the sample or fix on top of the sample. However, the rectification effect is much more efficient when the RF signal is directly applied to the sample via an electrode. The RF frequency is generated by a Rigol RSA5065N spectrum analyzer.

\textbf{Model simulations:} 

We present a detailed description of the theoretical modeling of the non-equilibrium transport on the surface of inhomogeneous topological surface states. 
The continuum single-electronic Hamiltonian describing the topological surface state is given as follows: 
\begin{equation}
H_{TSS}(\br)=v_F(\sigma_x p_y - \sigma_y p_x ) + V(\br) \sigma_0 + m(\br)\sigma_z,
\end{equation}
where $v_F$ is the Dirac velocity, $\boldsymbol{p}\equiv-i\hbar\nabla$ is the momentum operator, and $\left(\sigma_x,\sigma_y,\sigma_z\right)$ are Pauli matrices acting in the spin space. Whether it is the physical electronic spin or the pseudo-spin degree of freedom is not important for the purpose of this simulation. $m(\br)$ is a spatially varying mass term that opens an energy gap to the Dirac dispersion. $V(\br)$ is a spatially varying potential that is used as a tuning knob for mirror symmetry breaking. 

The electrodes are assumed to be described by a simple parabolic dispersion, 
$$H_{electrode}\left(\boldsymbol{r}\right)=\left(\frac{\boldsymbol{p}^2}{2m}-\mu\right)\sigma_0 ,$$ 
here $\mu$  tunes the Fermi level between the electrode relative to the topological surface states.  

To simulate the above ``electrode–topological surface state-electrode" set-up in the presence of an externally applied alternating electrical field, we use the open source software Tkwant. The topological surface state Hamiltonian is approximated by a hopping model on a square lattice, and given by (in second quantized form)
\begin{equation}
\begin{split}
    & H_{TSS}^{lattice} = \sum_{\br \in TSS} \sum_{\alpha\beta} \\
    & \left\{ \left[ i t \left( c_{\br,\alpha}^\dagger \sigma_{y,\alpha\beta} c_{\br+\hat{x},\beta} - c_{\br,\alpha}^\dagger \sigma_{x,\alpha\beta} c_{\br+\hat{y},\beta} \right) +h.c. \right] \right. \\
    & \left. + \left[ c_{\br,\alpha}^\dag\left(V(\br)\sigma_{0,\alpha\beta}+m(\br)\sigma_{z,\alpha\beta}\right)c_{\br,\beta} \right] \right\}.
\end{split}
\end{equation}
Here, $\hat{x}$ and $\hat{y}$ are used to denote a lattice constant along the x and y directions. It is important to note that these are not atomic lattice constants of \SM. As we study the continuum model, there is no length scale associated with theoretical modeling. In experiments, however, a length scale can be associated with the mean free path of topological surface states. Due to the suppressed backscattering, the mean free path can be quite long.

The electrode Hamiltonian is given by
\begin{equation}
\begin{split}
    & H_{electrode}^{lattice}  = \sum_{\br\in electrode}\sum_{\alpha\beta} \\
    & \left\{-t \left[ \left( d_{\br,\alpha}^\dagger\sigma_{0,\alpha\beta}d_{\br+\hat{x},\beta} + d_{\br,\alpha}^\dag\sigma_{0,\alpha\beta}d_{\br+\hat{y},\beta} \right)+h.c. \right] \right. \\
    & \left. -d_{\br,\alpha}^\dagger \left(\mu\sigma_{0,\alpha\beta}\right)d_{\br,\beta}\right\},
\end{split}
\end{equation}
where we choose the same hopping parameter $t$ as the lattice regularization of topological surface states. $\mu=-3.8t$ is the electrode chemical potential. Its value is unimportant, and chosen only to reduce the number of electronic states participating in the transport process, as the simulation can be computationally costly.

The tunneling Hamiltonian between the two electrodes (labeled as $\pm$) and the topological surface states is taken to be
\begin{equation}
\begin{split}
    & H_{tunneling,\pm}^{lattice}  = \sum_{\br\in interface} \sum_{\alpha\beta} \\
    & \left\{ -t\left[ \left(d_{\br,\alpha}^\dagger\sigma_{0,\alpha\beta}e^{i\phi_\pm(t)} c_{\br\pm\hat{x},\beta}  
    + d_{\br,\alpha}^\dagger \sigma_{0,\alpha\beta} d_{\br+\hat{y},\beta} \right) + h.c.\right] \right. \\
    & \left. -\left(d_{\boldsymbol{r},\alpha}^\dag\mu\sigma_{0,\alpha\beta}d_{\boldsymbol{r},\beta}\right)\right\},
\end{split}
\end{equation}
where $\phi_\pm\left(t\right)=\pm\phi_0\cos{\left(\omega_0t\right)}$ is the time-dependent Peierls phase describing an alternating electric field $E\left(t\right)=-\frac{d\phi\left(t\right)}{dt}$ at the two interfaces between electrodes and the topological surface states. 

The quantum transport simulation is computationally intensive. To conduct the simulation, we choose the topological surface state region to be a rectangle of length $L=20$ sites along the $x$-direction, and width $W=10$ sites along the $y$-direction. To study the effect of mirror symmetry breaking we used 
$$V\left(\boldsymbol{r}\right)=V_0\tanh{\frac{x-L/2}{\xi}}.$$
This breaks the mirror symmetry with respect to the $yz$-plane. To study the effect of an insulating region separating puddles of metallic topological surface states we used 
$$m\left(\boldsymbol{r}\right)=m_0\Theta\left(\xi_m-\left(x-\frac{L}{2}\right)\right)\Theta\left(x-\frac{L}{2}+\xi_m\right)$$
where $\Theta\left(x\right)$ is the unit step function. 
In Figure~\ref{fig4} of the manuscript, we used the following parameters: 
$$t=1,\ \phi_0=2,\ \omega_0=0.05,\ \xi=2,V_0=0.1,\ m_0=0,\xi_m=2.5.$$
Our results demonstrate current rectifying effects in the presence of mirror symmetry breaking, with and without an insulating gap separating the two regions of the topological surface states. In the Fig. S9, we show results for a few other choices of parameters, including when $m_0\neq0$.

\textbf{Acknowledgments} 

We thank Vidya Madhavan, Binghai Yan, Yuanfeng Xu, Yongqing Li, Joe Thompson, Cagliyan Kurdak, Wenxin Ding, Chunyu Guo, and Jinglei Zhang for fruitful discussions. We appreciate the technical support from Da Xu on the RF signal measurements and Chao Cao on numerical calculations. This study was funded in part by the National Key R\&D Program of China (Grant Nos. 2022YFA1402200 and 2023YFA1406100), the National Natural Science Foundation of China (Grant Nos. 12374151, 12350710785 12222410 and 12034017), the Zhejiang Provincial Natural Science Foundation of China (Grant No. LR25A040003), and the Fundamental Research Funds for the Central Universities (Grants Nos. 226-2024-00039 and 226-2024-00068). Work at Florida State University is supported by the NSF via Grants DMR-1905843 and DMR-2325147. A portion of this work was performed at the National High Magnetic Field Laboratory, which is supported by the NSF Cooperative Agreement (Nos. DMR-2128556 and DMR-1644779) and the State of Florida. Work at Los Alamos National Laboratory was performed under the auspices of the U.S. Department of Energy, Office of Basic Energy Sciences, Division of Materials Science and Engineering. Scanning electron microscope and energy dispersive X-ray measurements were supported by the Center for Integrated Nanotechnologies, an Office of Science User Facility operated for the U.S. Department of Energy Office of Science.

\textbf{Author contributions} 

L.J., X.W., and H.Y. conceived the project. The single crystals were provided by J.Z, S.T., and P.R., L.J., Z.H., J.Z., C.W., D.G., and S.T. conducted the measurements. X.W. and X.D. performed the theoretical analysis. L.J., Z.H., M.S., S.W., P.R., P.X., H.Y., and X.W. wrote the paper with input from all authors.


%

\end{document}